\documentclass{aa}  

\usepackage{graphicx}
\usepackage{natbib}
\usepackage{caption}
\usepackage{subcaption}
\usepackage{txfonts}
\begin{document}

   \title{Chromospheric seismology above sunspot umbrae}


   \author{B. Snow
          \and
          G. J. J. Botha
          \and
          S. R\'egnier
          }

   \institute{Department of Mathematics and Information Sciences, Northumbria University, Newcastle-Upon-Tyne, NE1 8ST, UK\\
              \email{ben.snow@northumbria.ac.uk}
             }


 
  \abstract
   {The acoustic resonator is an important model for explaining the three-minute oscillations in the chromosphere above sunspot umbrae. The steep temperature gradients at the photosphere and transition region provide the cavity for the acoustic resonator, which allows waves to be both partially transmitted and partially reflected.}
   {In this paper, a new method of estimating the size and temperature profile of the chromospheric cavity above a sunspot umbra is developed. }
   {The magnetic field above umbrae is modelled numerically in 1.5D with slow magnetoacoustic wave trains travelling along magnetic fieldlines. Resonances are driven by applying the random noise of three different colours---white, pink and brown---as small velocity perturbations to the upper convection zone. Energy escapes the resonating cavity and generates wave trains moving into the corona. Line of sight (LOS) integration is also performed to determine the observable spectra through SDO/AIA.}
{The numerical results show that the gradient of the coronal spectra is directly correlated with the chromosperic temperature configuration. As the chromospheric cavity size increases, the spectral gradient becomes shallower. When LOS integrations is performed, the resulting spectra demonstrate a broadband of excited frequencies that is correlated with the chromospheric cavity size. The broadband of excited frequencies becomes narrower as the chromospheric cavity size increases.}
   {These two results provide a potentially useful diagnostic for the chromospheric temperature profile by considering coronal velocity oscillations.}


   \maketitle
%

\section{Introduction}

There are three main categories for the oscillations that occur in and around sunspots: five-minute oscillations at photospheric levels, three-minute oscillations above the umbra, and running waves moving away from the umbra along penumbral structures \citep{Bogdan2000,Bogdan2006}. Many observations show the existence of three-minute waves. \cite{Marsh2006} observed these oscillations through both the 171 \AA~ TRACE and the SOHO/CDS instruments. The three-minute oscillations have also been observed at the transition region in the microwave band as a modulation of gyroresonant emission \citep{Shibasaki2001} and in ultraviolet wavelengths \citep{demoortel2002}. \cite{Thomas1987} observed multiple peaks in the three-minute band (4.5-10 mHz) from both space and ground instruments. A similar excited range of frequencies is present in \cite{Rez2012}. A recent overview of magnetohydrodynamic waves and coronal seismology can be found in \cite{DeMoortel2012}. 

There are three competing mechanisms to explain the source of the three-minute oscillations: excitation of the cut-off frequency, the wake of propagating shock waves, and a chromospheric resonator. \cite{Fleck1991} suggested that the three-minute oscillations are due to a basic physical effect: the excitation of waves at the cut-off frequency. They numerically modelled a piston driving waves at the photosphere and a simplified isothermal atmosphere that ignored many complexities including temperature gradients and non-linearities. The model produced three-minute oscillations. \cite{Fleck1991} also investigated a VALC profile with similar results. Their explanation was that waves above the cut-off frequency are free to propagate upwards, increasing in amplitude, whereas waves below the cut-off frequency are damped. This is however, insufficient to explain the observed amplification of the three-minute oscillations \citep{Stangalini2012}.

It has also been suggested that the three-minute oscillations arise from the wake of propagating shock waves, \citep[see][]{Hansteen1997}. These shocks are also present in the 1D numerical simulations of \cite{Bard2010}. However, 1D simulations of propagating shock waves leads to unrealistic shock merging \citep{Ulmschneider2005}.

An alternate explanation for the existence of three-minute oscillations above sunspot umbrae is the presence of an acoustic chromospheric resonator. Above the chromosphere in the transition region and below it in the photosphere, there are large temperature gradients that provide the semi-permeable boundaries necessary for resonances of slow magnetoacoustic waves to occur inside the chromosphere \citep{Zhugzhda2008}. When a wave encounters these large temperature gradients, part of the wave is transmitted and part is reflected. We note that the model set-up in the previously mentioned models \citep{Bard2010,Fleck1991} prohibits any chromospheric resonator since the waves are driven using a piston at the photosphere. \cite{Zhugzhda2008} also investigated how frequencies above and below the cut-off frequency behave in this resonator by using a temperature that changes non-monotonically with height. Waves above the cut-off frequency are partially reflected due to the sharp temperature gradients at the photosphere and transition region in the sunspot atmosphere. Waves below the acoustic cut-off frequency cause the chromosphere to resonate between the transition region and the point at which the cut-off frequency drops below the wave frequency \citep{Taroyan2008}. The resonances cause oscillations that leak energy through the transition region into the corona. This causes regular waves to travel into the corona. \cite{Botha2011} performed numerical simulations of a chromospheric resonator with a perturbation in the form of a single pulse. Two such pulses were considered, two and five minutes, propagating along a magnetic field line above a sunspot umbra. The results numerically proved the existence of an acoustic resonator and the production of the familiar three-minute oscillations.

Observational results show that the amplitude of the three minute oscillation is not explained well enough by the excitation of the cut-off frequency \citep{Stangalini2012}. Furthermore, many observations show a harmonic wave structure \citep{Yuan2011, Rez2012, Sych2011} as opposed to the saw-tooth signal required for the shock wave model. As such, the chromospheric resonator model is a promising explanation for the three-minute oscillations.

In this paper, perturbation noise is applied in the upper convection zone and allowed to propagate along the magnetic field line above a sunspot umbra. The wave trains will be interpreted as slow magnetoacoustic waves propagating with the local sound speed \citep{Roberts2006}. The applied noise is of the form $1/f^{\xi}$, corresponding to solar granulation noise \citep{RS1997}. 

This paper investigates the effect of different stochastic noise signatures on the chromospheric resonator. The effect of different chromospheric temperature configurations on the resulting coronal spectra is considered (Section \ref{sec_res}). The findings indicate a novel method for estimating the chromospheric temperature configuration based upon the frequency spectra present in the wave trains that escape from the resonating chromosphere into the corona. Line-of-sight (LOS) integration is then performed on the data using several SDO/AIA response functions (Section \ref{sec_LOS}). 

\section{Model}

A 1.5D numerical investigation was performed to simulate the effects of different noise signatures and chromospheric temperature configurations on the coronal frequency spectra. This involves the generation of noise, which is described in Section \ref{sec_noise}. Summaries of the numerical configuration and the post-processing method are also provided.  

\subsection{Numerical set-up}

The ideal nonlinear compressible magnetohydrodynamic (MHD) equations are implemented numerically using Lare2D, as presented by \cite{Arber2001}. These equations are as follows:

\begin{eqnarray}
\frac{\partial \rho}{\partial t} &=& - \nabla \cdot (\rho \textbf{v}) \\
\frac{\partial \textbf{v}}{\partial t} + \textbf{v} \cdot \nabla \textbf{v} &=& \frac{1}{\rho} \textbf{J} \times \textbf{B} - \frac{1}{\rho} \nabla P + \textbf{g}\\
\frac{\partial \textbf{B}}{\partial t} &=& - \nabla \times \textbf{E} \\
\frac{\partial \epsilon}{\partial t} + \textbf{v} \cdot \nabla \epsilon &=& - \frac{P}{\rho} \nabla \cdot \textbf{v}\\
\textbf{E} +\textbf{v} \times \textbf{B} &=& 0 \\
\nabla \times \textbf{B} &=& \mu _{0} \textbf{J} \\
P &=& \frac{\rho k_B T}{\mu _m} \\
\epsilon &=& \frac{P}{\rho (\gamma -1)} \label{eqn_internalenergy}
\end{eqnarray} 
where $\rho$, $P$, $T$, and $\textbf{v}$ are the plasma density, pressure, temperature, and velocity, respectively. Here, $\textbf{J}$ is current density, $\textbf{B}$ is magnetic field, $\textbf{E}$ is electric field, and $g$ is gravity. The internal energy $\epsilon$ is given by Equation (\ref{eqn_internalenergy}). Here, $\mu _m = 1.4 m_p$  where $m_p$ is the mass of a proton, $k_B$ is the Boltzmann constant, $\mu _0$ is the permeability of free space, and $\gamma =5/3$ is the ratio of specific heats.

The normalisation values for length, magnetic field, and density are $L_0 =150 \times 10^{3}$ m, $B_0 = 0.12$ T, and $\rho _0 = 2.7 \times 10^{-4}$ kg m$^{-3}$. Gravity is constant as 274 m s$^{-2}$ for the majority of the domain, however above approximately 44 Mm the gravity is gradually reduced to zero on the upper boundary. This is to allow for the specification of an open boundary and prevent reflections from the upper boundary.

The velocities along a unidirectional magnetic field line above the umbra, perpendicular to the photospheric plane, are considered. The magnetic field acts as a waveguide for slow magnetoacoustic waves propagating into the corona. These slow waves propagate parallel to the magnetic field at the acoustic speed, which is independent of magnetic field strength. This allows us to choose in our 1.5D model a constant magnetic field strength of 10 Gauss, in line with the chromospheric strength of a sunspot umbral magnetic field. 

The full equations are solved in the three Cartesian coordinates ($x,y,z$) with $y$ taken as the vertical direction. The numerical grid contains 1024 cells in this direction. The two horizontal directions ($x,z$) are invariant, making the simulation 1.5D. There are four cells in the invariant $x$ direction. A convergence test of doubling the grid resolution in $y$ resulted in no new features hence 1024 cells are deemed sufficient. The grid spans the range -68 Mm to 68 Mm, leading to a vertical cell size of 132,927 m. The chromosphere is positioned in the middle of the computational domain. The large domain size is used to prevent boundaries from interfering with the resonator. The end time is set as $10^{4}$ seconds.

The numerical dissipation in the code was tested by launching a slow pulse in a uniform plasma \citep[e.g.][]{Yuan2015} using a range of values: $\rho = 10^{-5}, 10^{-11},$ and $10^{-12}$ kg m$^{-3}$ and $T = 0.4$ MK, and $1$ MK. The maximum amplitude and full width at half maximum were consistent to four and seven significant figures, respectively when calculated at the start and end of the domain. Therefore the numerical algorithm used here is not significantly dispersive.

\subsection{Temperature profiles}

\begin{figure}
\centering
\includegraphics[width=88mm]{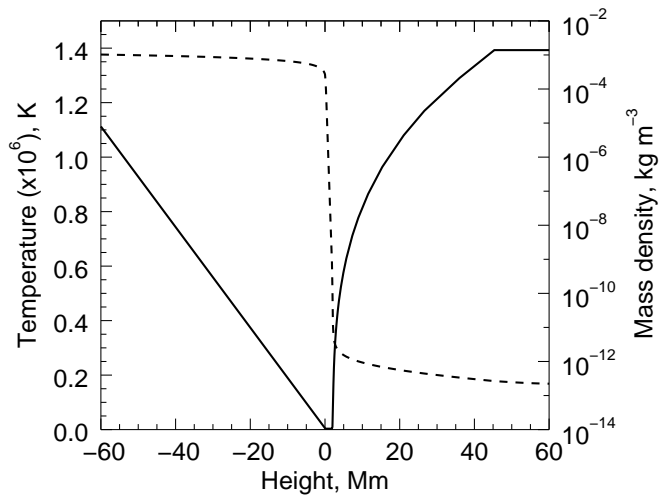}
\caption{Initial temperature (solid line) and density (dashed line) profiles used in the simulation. The photosphere is located at height 0.}
\label{fig_temp}
\end{figure}

The coronal temperature profile is obtained from the model by \cite{Avrett2008}, with the chromospheric temperature being replaced by the investigated profiles. To prevent velocity perturbations being reflected from the upper boundary, it is necessary that the temperature gradient at the top of the computational domain is zero. Therefore, from approximately 44 Mm above the photosphere and higher the temperature is flattened. This has no effect on the simulation results and the data are investigated at points below this region. The temperature profile is shown in Figure \protect\ref{fig_temp}. 

For the convection zone, a polytrope temperature profile is used. The lower velocity boundary is specified directly from the input noise seed. By setting the lower boundary far away from the resonator, downwards propagating waves are damped and do not reflect from the lower boundary. 

\begin{figure*}
\centering
\begin{subfigure}[t]{0.45 \textwidth}
\includegraphics[width=82mm]{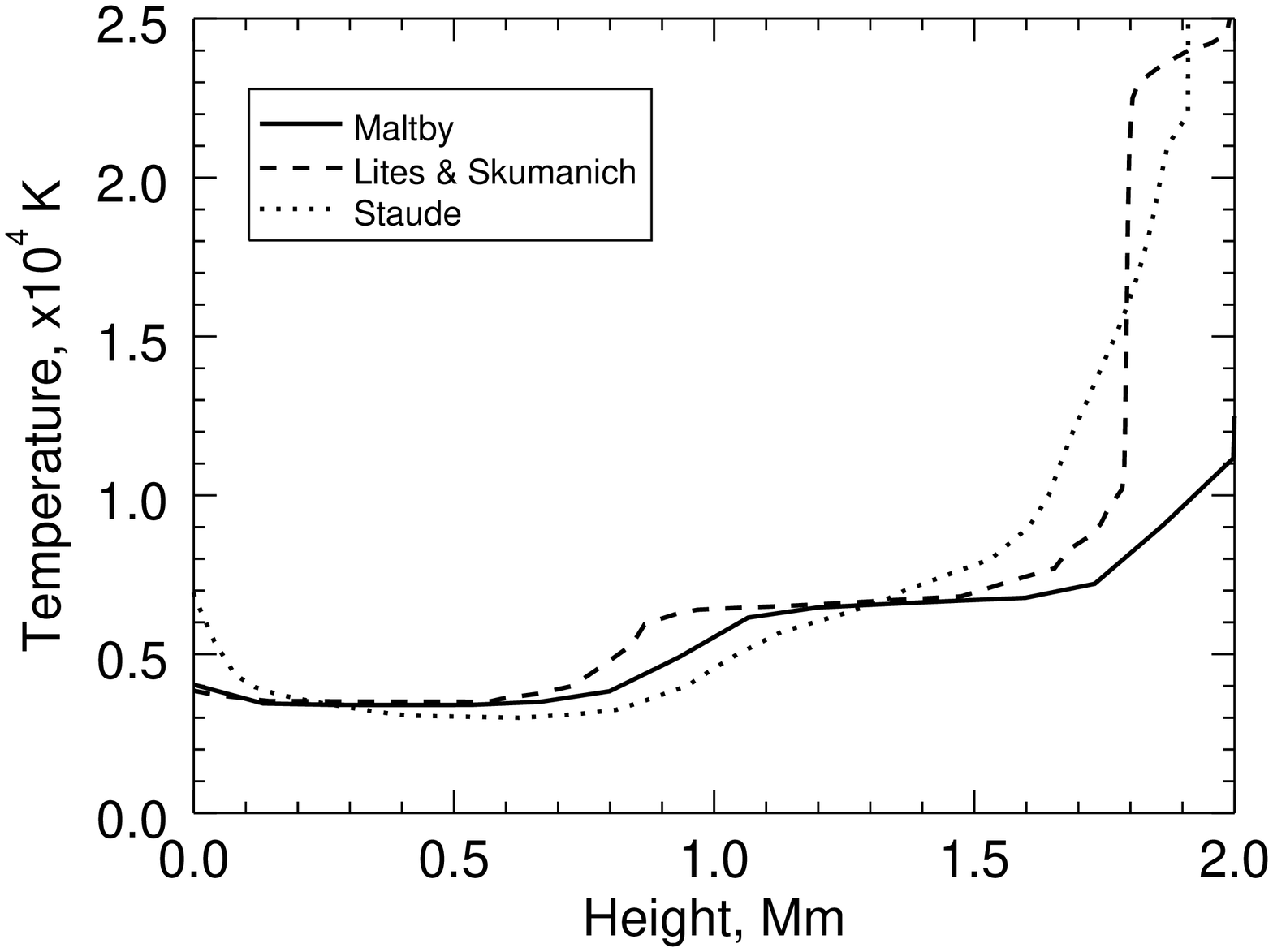}
\caption{}
\label{fig_theoret}
\end{subfigure}%
\hspace{1cm}
\begin{subfigure}[t]{0.45 \textwidth}
\includegraphics[width=82mm]{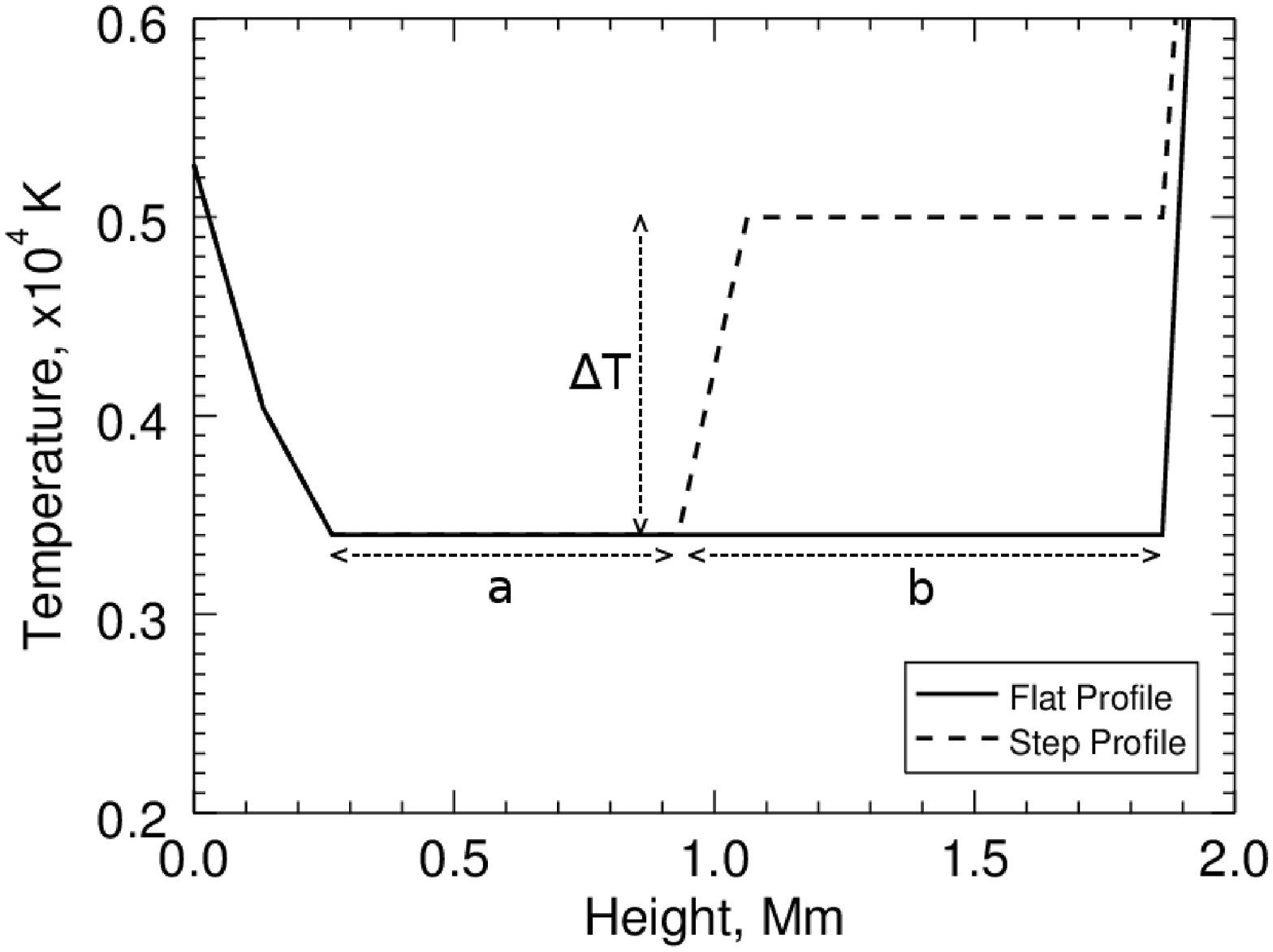}
\caption{}
\label{fig_profiles}
\end{subfigure}
\caption{Chromospheric temperature profiles. (a) Different theoretical temperature profiles for the chromosphere above a sunspot: solid line from \cite{Maltby1986}, dashed line from \cite{Lites1982}, and dotted line from \cite{Staude1981}. (b) Flat and step chromospheric temperature profiles with the photosphere at 0 and the transition region at 1.8 Mm. This chromospheric profile is shown in relation to the photospheric and coronal temperature profiles in Figure \protect\ref{fig_temp}. The step profile is a simplification of the theoretical temperature models shown in Figure \protect\ref{fig_theoret}.}
\end{figure*}

There are several theoretical models available for the structure of the chromosphere above a sunspot \citep{Staude1981, Lites1982, Maltby1986} as shown in Figure \ref{fig_theoret}. These theoretical models suggest that the chromospheric cavity size is between 1.5 and 1.8 Mm. Here two profiles are investigated, a flat and a step profile (Figure \ref{fig_profiles}) and they have a chromospheric cavity size $a + b$. In the numerical model, $a$ is the size of the lower chromospheric temperature plateau and $b$ is the size of the upper chromospheric temperature plateau. The flat profile allows for consideration of an idealised scenario, whereas the step profile is closer to the theoretical models with a lower and an upper chromospheric temperature plateau, as well as a temperature jump at the mid-chromosphere.

At both the photosphere and transition regions there is a steep temperature gradient, as can be seen in Figures \protect\ref{fig_temp},  \ref{fig_theoret} and \ref{fig_profiles}. These act as semi-permeable boundaries that allow energy to be partially reflected and partially transmitted. The energy partially reflected back into the chromosphere results in resonance in the chromospheric cavity that exists between the photosphere and transition region \citep{Zhugzhda2008,Botha2011}.

In \cite{Botha2011}, where a single pulse was used to energise the chromospheric cavity, the resonator found its natural frequencies. In the present study, a continuous noisy source of energy enters the chromospheric cavity. This results in a broad energisation of a range of frequencies, as will be discussed in Section \ref{sec_freqreg}. In addition, energisation is caused by the partial reflection and partial transmission of the energy content at the photosphere and at the transition region.

Multiple simulations were performed, varying the chromospheric cavity size and velocity source. This changes the energy content and energy distribution of the oscillations present in the resonator.

\subsection{Velocity initialisation}
\label{sec_noise}
The resonator will be perturbed by applying random noise in the upper convection zone, 68 Mm below the photosphere. Granulation noise follows a $1/f^\xi$ distribution where the power is proportional to the reciprocal of the frequency \citep{RS1997}. Three types of noise will be considered; white (uniform distribution), pink ($1/f$) and brown ($1/f^2$). This allows investigation into the effect of different perturbations. The velocity signal applied at the source is therefore of the form 

\begin{equation}
v(t) = \sum ^\infty _{n=0} a_n \sin (n \pi t + p_n)
\end{equation}
where the constants $a_n$ is determined by the noise colour and $p_n$ is a random phase shift.

\begin{figure}
\includegraphics[width=82mm]{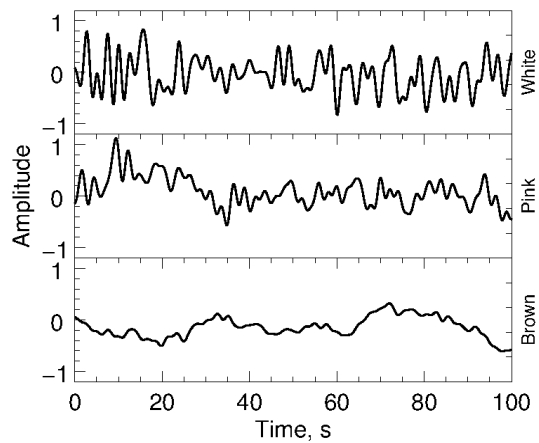}
\caption{Example time series of the different noise sources, which are applied to the lower boundary as slow magnetoacoustic velocity perturbations.}
\label{fig_noisesource}
\end{figure}

White noise is the standard random noise where there is equal power in all frequencies. However since both pink and brown noise decline in power as the frequencies increase, these two are biased towards lower frequencies. Examples of the three types of noise are shown in Figure \ref{fig_noisesource}.

\begin{figure}
\includegraphics[width=82mm]{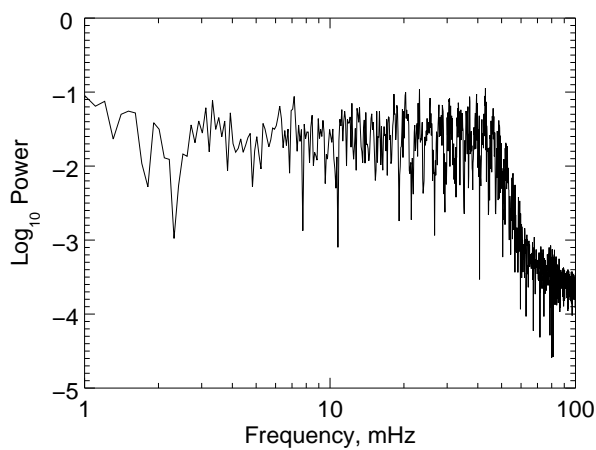}
\caption{Frequency spectrum of the white noise sample shown in Figure \protect\ref{fig_noisesource}. The pink and brown noise show similar spectra, with gradients of -1 and -2 respectively.}
\label{fig_whitespec}
\end{figure}

The noise samples will be generated using a filter method. This involves generating a white noise sample. The white noise sample is then converted to spectral space via a fast Fourier transform (FFT). The spectra are multiplied by the desired frequency distribution and then the signal is converted back to physical space with an inverse FFT. Radial basis functions \citep{Buhmann2000} are then used to smooth the signal. An FFT is then performed again to this smoothed signal and the gradient is calculated to ensure that it is within acceptable bounds ($\pm 0.1$) of the desired gradient. Due to the stochastic nature of noise, multiple samples of each colour were tested to ensure reliable conclusions can be deduced. To have a fully resolved spectra would require infinite random values to be generated at the initial stage. However, this is not possible. Here 1024 random numbers are used and the signal is well resolved up to approximately 50 mHz. A signal resolved to a higher frequency, 100 mHz, was tested. There was no difference in the results and the lower cut-off is used throughout this paper, i.e. 50 mHz. An example of the frequency spectrum for white noise is given in Figure \ref{fig_whitespec}. The velocity is scaled to be between $\pm 1$ ms$^{-1}$ at the lower boundary, 68 Mm below the photosphere, allowing for the consideration of small perturbations.

\begin{figure}
\centering
\includegraphics[width=82mm]{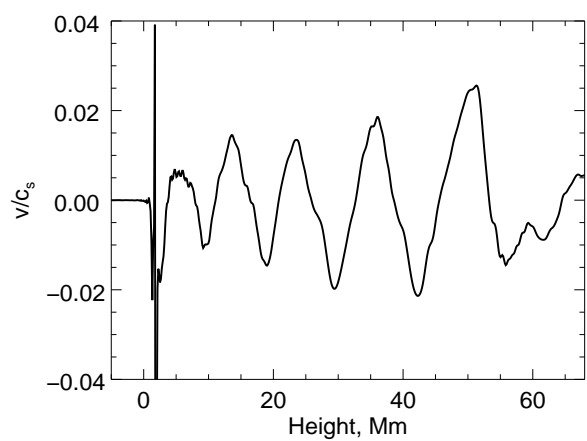}
\caption{Snapshot of the evolution of velocity across the domain at one time instance. The source is located at $-68$ Mm and the photosphere is at $0$ Mm. Large amplitude velocity is present in the chromosphere due to the high energy content of the resonating cavity. Wave trains propagating into the corona are reasonably well structured. Time evolution of velocity at different points is shown in Figure \protect\ref{fig_vel4}. For this plot a flat chromospheric temperature profile of depth 1.8 Mm was used with a white noise velocity source.}
\label{fig_vel_domain}
\end{figure}

The applied velocity corresponds to granulation noise. However the source is far below the height at which one would expect granules to form. The justification for this is that the nature of the upwards propagating velocity perturbation from the source does not change in the model's convection zone, i.e. the distribution of noise being fed into the chromospheric resonator is the same as the distribution at the source, only the amplitude changes. The lower boundary is set far from the photosphere (-68 Mm) to allow the downwards propagating waves from the resonator to disperse naturally without reflection from the lower boundary.  

\begin{figure}
\includegraphics[width=82mm]{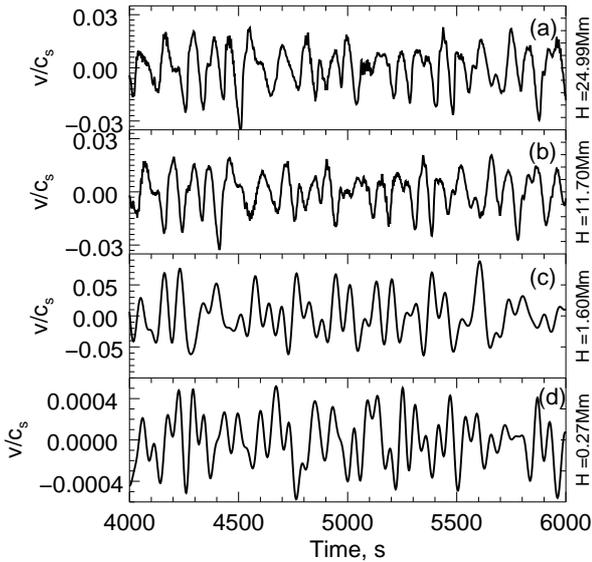}
\caption{Plots of velocity/sound speed vs time for different heights H above the photosphere. (a) and (b) are in the corona, (c) is located in the upper chromosphere, (d) in the lower chromosphere. The model set up for this plot was a flat chromospheric temperature profile of depth 1.8 Mm with a white noise velocity source.}
\label{fig_vel4}
\end{figure}

\section{Transmission into the corona}
\label{sec_res}

\subsection{Velocity propagation from the chromosphere to the corona}
\label{sec_velevo}
\begin{figure}
\includegraphics[width=82mm]{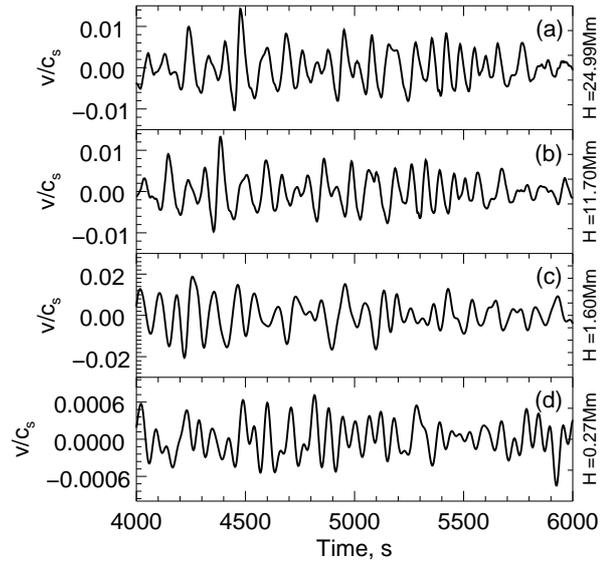}
\caption{Plots of velocity/sound speed vs time for different heights H above the photosphere. (a), (b), (c) and (d) are at the same locations as in Figure \protect\ref{fig_vel4}. A step temperature profile was used with $a=b=0.9$ Mm for both the upper and lower plateaux. The perturbation noise was white.}
\label{fig_vel4step}
\end{figure}

The behaviour of the velocity perturbations at different points in the solar atmosphere is shown in Figures \ref{fig_vel_domain}, \ref{fig_vel4} and \ref{fig_vel4step}. Figure \ref{fig_vel_domain} shows a snapshot of the velocity across the domain. Below the photosphere the velocity is too small to be seen on the plot. However there are wave trains propagating both towards the resonator, from the velocity source applied at the lower boundary, and away from the resonator towards the lower boundary, due to downwards partially transmitted velocity perturbations. 

A time series of the velocity perturbations measured at a point in the lower and upper chromosphere is shown in Figures \ref{fig_vel4} and \ref{fig_vel4step}. The wave trains propagating into the corona appear well structured, i.e. well resolved with no shocks forming, with several higher frequencies present. This is reflected in the frequency spectra. \\ Figure \ref{fig_vel4} shows the time evolution of the velocity at different spatial points. The velocity amplitude, with respect to the local sound speed, increases throughout the chromosphere. When energy is transmitted through the temperature boundary at the transition region, there is a decrease in power since the energy is partially reflected back into the chromosphere. Again the signal appears to be mostly regular with several higher frequencies present, identifiable from the small scale fluctuations. Note that the two coronal figures, (a) and (b) in Figure \ref{fig_vel4}, show the same velocity perturbation shifted by time. This shows that once the signal has entered the corona, there are no significant changes to the velocity perturbation. There is no evidence of any shocks occurring and the signal is well resolved. The frequency regimes discussed in Section \ref{sec_freqreg} are all present when sampled at various points in and above the transition region.

\subsection{Frequency analysis}

\label{sec_nyquist}

An FFT was performed to analyse the frequencies present in the corona for each simulation. This involved considering the time series velocity fluctuations through a single point in space. A line has then been fitted to the FFT to quantify the relation between power and frequency using the well established method of least squares. The gradient of this line is then used to consider the effect the chromospheric temperature configuration has on the signal propagating through the corona.   
Velocity data are sampled at intervals of 0.25 seconds, leading to a Nyquist frequency of 2000 mHz.

\subsection{Coronal frequency regimes}
\label{sec_freqreg}

\begin{figure}
\centering
\includegraphics[width=88mm]{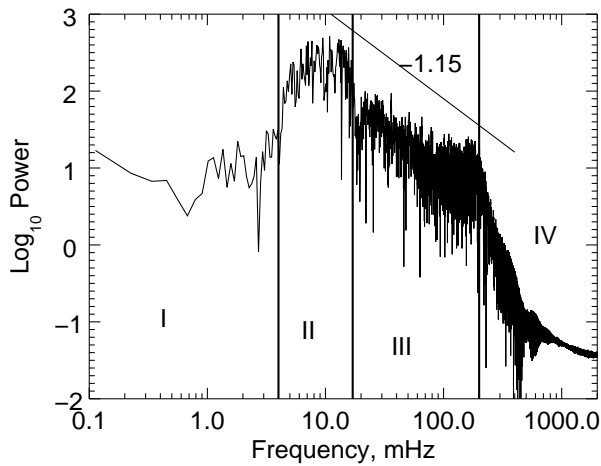}
\caption{Four frequency regimes present in the spectra. I is the lower cut-off due to the acoustic cut-off frequency (below 4 mHz). II is the broad peak (4-17 mHz). III is the gradient decline (17-200 mHz). IV is the upper cut-off (above 200 mHz). This spectrum was generated for a flat chromospheric profile of depth 1.8 Mm with a white noise velocity source. The line indicates the gradient of the power in region III, which is -1.15 in this case.}
\label{fig_freq_reg}
\end{figure}

\begin{figure}
\centering
\includegraphics[width=88mm]{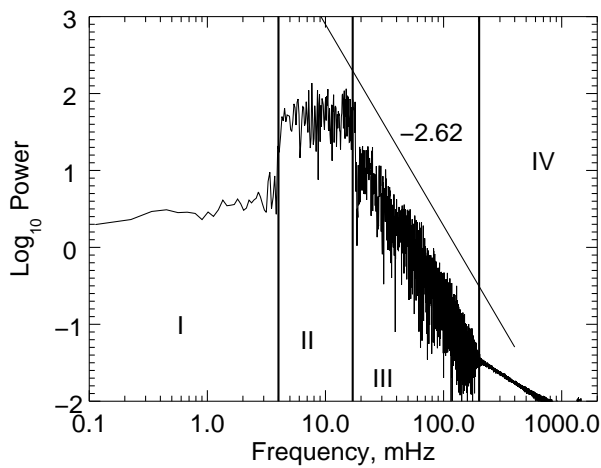}
\caption{Four frequency regimes present in the spectra. Regions I, II, III and IV are the same as in Figure \protect\ref{fig_freq_reg}, generated for a step chromospheric profile with upper and lower plateau depths of $a=b=0.9$ Mm and a white noise velocity source. $\Delta T$ is 2600 K. The line indicates the gradient of the power in region III, which is -2.62 in this case.}
\label{fig_freq_reg_step}
\end{figure}

The frequency for each simulation is sampled through time in the corona at a point 25 Mm above the start of the transition region. Four main regions are observed in the spectra for all the simulations, as shown in Figures \protect\ref{fig_freq_reg} and \ref{fig_freq_reg_step} for the flat and step profiles respectively; all except the lower cut-off frequency are due to the resonating behaviour. There is a general trend of increased spectral power for larger chromospheres. This is expected since there is a larger region for the resonances to oscillate in, which provides increased amplitude and hence higher power. Note also that there is a level of uncertainty present in the data due to the fact that the system is being driven by stochastic noise. This arises because the noise is generated using a finite number of samples resulting in slight deviations from the expected power at each frequency. At double the grid resolution, the four regimes were still present at the same frequencies.

\paragraph{\textbf{Region I: Lower Cut-off}}

Below approximately 4 mHz there is a very low power region (Figure \ref{fig_freq_reg}, region I). This is due to the acoustic cut-off frequency $\omega _c$ that allows waves to propagate upwards with frequencies

\begin{equation}
\omega > \omega _c = \frac{c_s}{2H} \sqrt{1+2 \frac{\partial H}{\partial z}} 
\end{equation}
where $H = c_s^2/(\gamma g)$, $\gamma$ is the ratio of specific heats, $c_s$ is the sound speed, and $z$ is the height \citep{Roberts2004}. Typical values of the acoustic cut-off frequency are in the range 3.5 - 5.2 mHz \citep[e.g.,][]{Bel1977, Mcintosh2006, Yuan2014}.

\paragraph{\textbf{Region II: Broad Peak}}

Between approximately 4 and 17 mHz the power is relatively flat (Figure \ref{fig_freq_reg}, region II). This corresponds to a broad band of frequencies with oscillation times of between approximately one-minute to four-minutes. Following the work of \cite{beckers1972oscillatory}, several papers have investigated the three-minute oscillations above sunspot umbra. For example, \cite{Rez2012} present spectra for oscillations above sunspot umbrae through various filters which show an excited range of frequencies between approximately 6-8 mHz. \cite{Thomas1987} makes reference to a three-minute band of frequencies between 4.5 and 10 mHz containing multiple peaks.  

This broad peak is present in the same frequency range regardless of temperature configuration or perturbation noise. No discrete harmonics are visible in the spectra. This does not mean that the harmonics are not present; they are not visible due to the power distribution in the frequency range where the harmonics are situated. 

\paragraph{\textbf{Region III: Gradient Decline}}

The next region (Figure \ref{fig_freq_reg}, region III) between 17 and 200 mHz, is a region where the power is proportional to $1/f^\alpha$, which corresponds to a linear decline of gradient $\alpha$ when plotted on log-log axes. This region is found to vary in gradient when the temperature profile is modified and provides the diagnostic tool for the main conclusions drawn in this paper. This is discussed in more detail in the Sections \ref{sec_flat} and \ref{sec_step}. Confidence intervals were calculated to ensure the fitted line was representative of the data. The standard deviation was small in all cases, ranging from 0.041 to 0.052, resulting in very narrow confidence intervals, identical to 3 decimal places. For one such line, the 95\% confidence interval is $-0.9649$ to $-0.9646$, with the calculated value being $-0.9647$.
 
\paragraph{\textbf{Region IV: Upper Cut-off}}

Above approximately 200 mHz (Figure \ref{fig_freq_reg}, region IV) the power drops significantly. The cut-off prevailed at the same frequency with a higher resolved input signal indicating that this upper cut-off is due to neither the input signal (Section \ref{sec_noise}) nor the sampling rate (Section \ref{sec_nyquist}). This is a fluid model which does not account for the ion-gyrofrequency. However the ion-gyration frequency is approximately $\Omega _i = 9.6 \times 10^4$ rad s$^{-1}$ \citep{priest2014magnetohydrodynamics}. This is far larger than the frequencies simulated in this model, therefore the obtained frequencies are not beyond the physical limitations of the fluid model. This suggests a physical phenomena for the upper cut-off unidentified at present. Due to the high frequency where the cut-off occurs, it cannot be resolved in observations at the present time. 
 
\subsection{Varying chromospheric cavity configuration} 
\subsubsection{Flat Profile}
\label{sec_flat}

\begin{figure*}
\centering
\begin{subfigure}[t]{0.45 \textwidth}
\includegraphics[width=80mm]{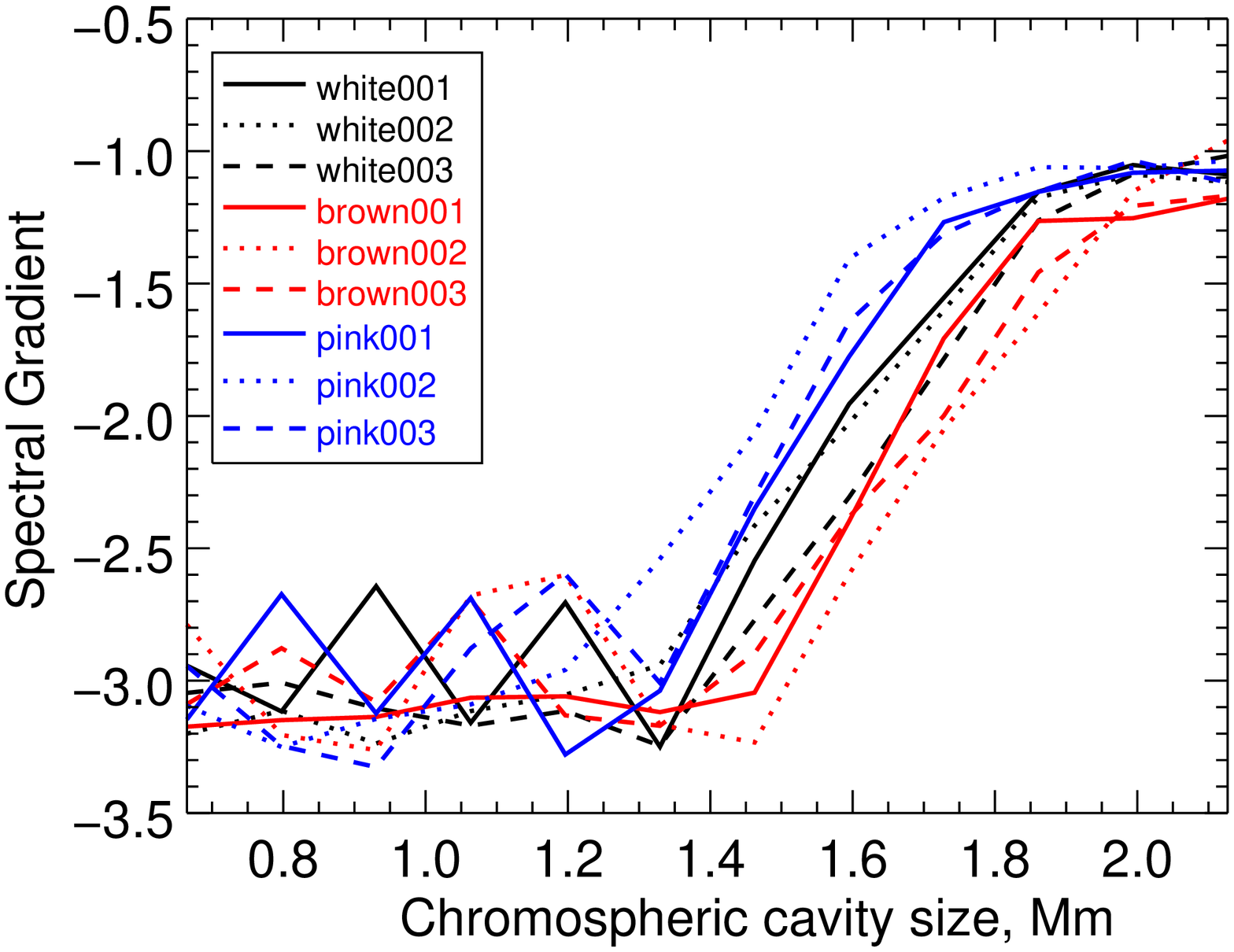}
\caption{}
\label{fig_flatgrads}
\end{subfigure}%
\hspace{1cm}
\begin{subfigure}[t]{0.45 \textwidth}
\includegraphics[width=80mm]{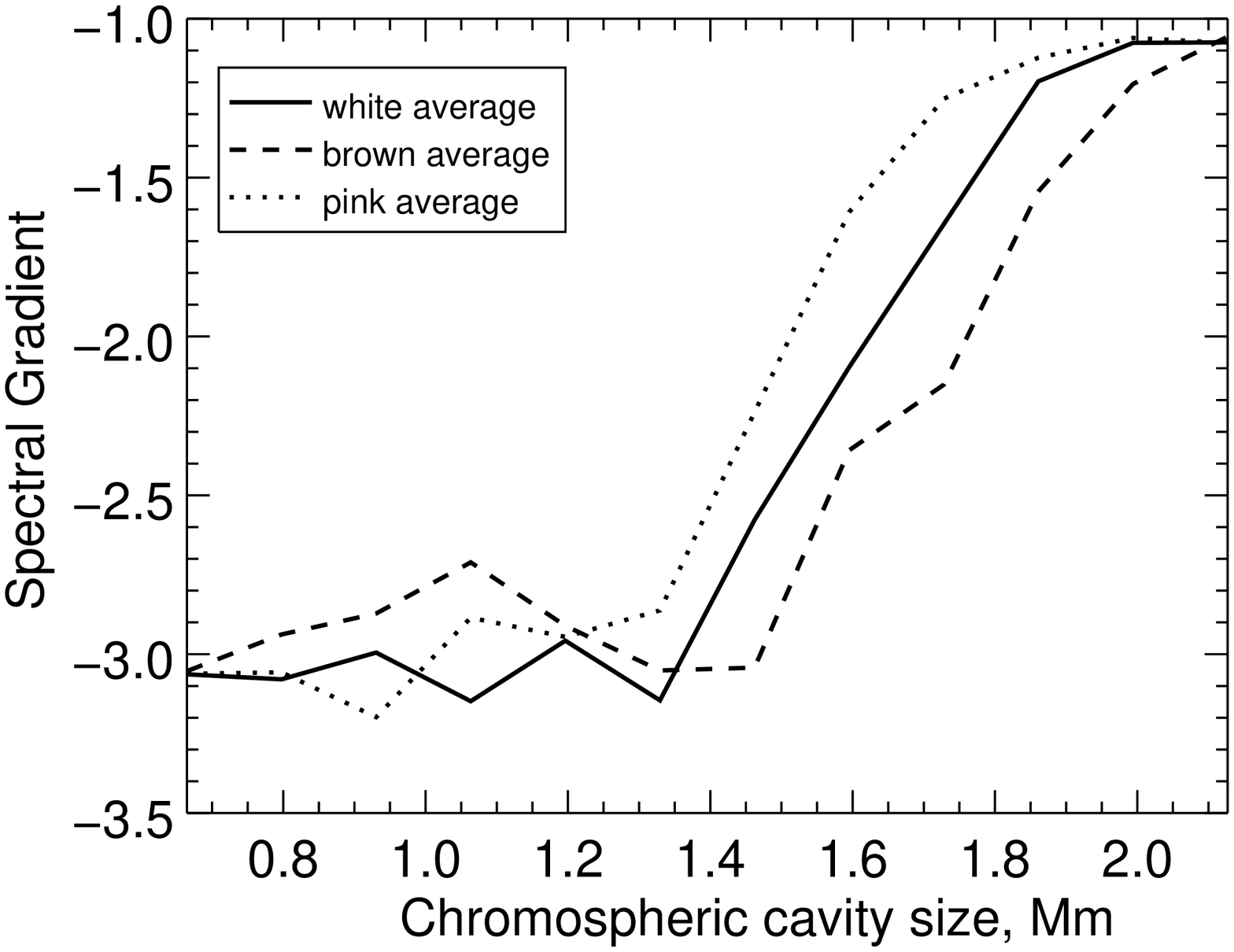}
\caption{}
\label{fig_flatgradsave}
\end{subfigure}
\caption{Gradients vs chromospheric cavity size for three samples of each noise colour (a), and averaged across each noise colour (b). The gradients were obtained in region III of Figure \protect\ref{fig_freq_reg}.}
\end{figure*}

Initially the temperature configuration of the chromosphere was assumed to be uniform, as shown in Figure \protect\ref{fig_profiles} with $\Delta T=0$. This was to investigate the behaviour in an idealised scenario. 
Here the parameter space consists of the chromospheric cavity size ($a+b$ in Figure \ref{fig_profiles}) and the perturbation noise colour. Three noise samples for each noise colour were used and a range of chromospheric cavity sizes between 0.66 and 2.13 Mm were tested. The gradient of the aforementioned frequency range (Figure \ref{fig_freq_reg}, region III) was calculated for each simulation. Note that due to the stochastic nature of noise, different samples of the same noise colour create slightly different results, indicated by the variations present in Figure \protect\ref{fig_flatgrads}. Multiple seeds were tested for this reason. 

For clarity, the gradients were averaged for each noise colour and are shown in Figure \protect\ref{fig_flatgradsave}.  This effectively removes some of the uncertainty present when considering noise. The full set of results is shown in Figure \protect\ref{fig_flatgrads}, showing the extent of randomness. The variation in the gradient is approximately $\pm0.5$ for narrow chromospheres of cavity size less than 1.3 Mm, whereas for large chromospheres of cavity size larger than 1.8 Mm there is virtually no variation in gradient across samples.

For narrow chromospheric cavities, less than 1.3 Mm, all three noise colours appear to have the same output gradient in their spectra, as shown in Figures \ref{fig_flatgrads} and \ref{fig_flatgradsave}. There is a sharp change in output gradient for chromospheric cavity sizes starting from around 1.3-1.4 Mm and ending at approximately 1.8 Mm. Here the gradient rapidly goes from steep, $\alpha \approx -3$, to relatively flat $\alpha \approx -1$. Since this large change in gradient is present, it may be possible to estimate the depth of the chromosphere from the spectra sampled in the corona at 25 Mm above the transition region.

The dependence of the power gradient on the chromospheric cavity size can be explained by considering the energy content of the chromosphere. As the chromospheric cavity size increases, the total energy content of the resonating waves increases. This allows for energy to be more evenly distributed among the higher frequencies when partially transmitted into the corona. For a thinner chromosphere, there is a low total energy content and as such only the lower frequencies become energised. As the chromosphere cavity increases in size, the energy content increases and there is more energy in the higher frequencies, relative to the lower frequencies, i.e. a decrease in the slope of the  power gradient in the FFT spectra (Figure \ref{fig_flatgradsave}).

With regards to noise, there does not appear to be a significant difference between noise colours. Whilst the location of the sharp gradient change in the spectra measured for narrow chromospheres appears slightly different in Figure \ref{fig_flatgradsave} for different noise colours, there is no distinct relationship and the variation in the location of the sharp change is small, occurring between approximately 1.3 and 1.4 Mm of chromospheric thickness. Therefore the discrepancy is most likely due to the stochasticity present.

For large chromosphere depths, above 1.8 Mm, the output gradient becomes fairly constant for all three noise colours, as shown in Figures \ref{fig_flatgrads} and \ref{fig_flatgradsave}

\subsubsection{Step profile}
\label{sec_step}

\begin{figure}
\centering
\includegraphics[width=88mm]{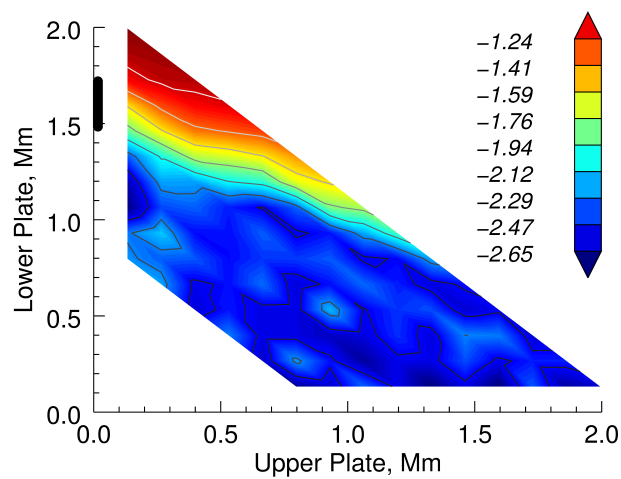}
\caption{Contours of the gradient in the obtained coronal spectra (region III of Figure \protect\ref{fig_freq_reg}) generated by white noise and the step temperature profile. The lower plateau sizes are given by values of $a$ and the upper plateau sizes by $b$ in Figure \protect\ref{fig_profiles}. The black marker on the vertical axis indicates the steep gradient change for the flat profile, as seen in Figure \protect\ref{fig_flatgradsave}.}
\label{fig_stepw}
\end{figure}

Next, the chromospheric temperature configuration is replaced with a step profile. This consists of two uniform temperature regions or plateaux, $a$ and $b$, separated by a small temperature jump, $\Delta T$ as shown in Figure \protect\ref{fig_profiles}. This is similar to theoretical chromospheric temperature models (Figure \ref{fig_theoret}) and provides a more realistic configuration. Thus the parameter space here consists of 4 variables: lower plateau size $a$, upper plateau size $b$, temperature jump $\Delta T$, and noise colours. In this section, the temperature jump will be fixed at 2600 K since this agrees with the theoretical models (Figure \ref{fig_theoret}). Varying the temperature jump will be investigated in Section \ref{sec_tempjump}.

The depth of both the lower ($a$) and upper ($b$) plateau varied in the range $0.13$ Mm $\leq a,b \leq 2.0$ Mm, under the constraint that the total depth $0.8$ Mm $\leq a+b \leq 2.13$ Mm. This provides a maximum total chromospheric cavity size of approximately the largest value from the theoretical models, i.e. 2 Mm as shown in Figure \ref{fig_theoret}. The minimum chromospheric cavity size is 0.8 Mm. The three types of noise were used as in Section \ref{sec_flat}, however on this occasion only one sample of each colour was used. The gradient of the coronal spectra was calculated as before in region III of Figure \ref{fig_freq_reg_step}. The results can be seen in Figure \protect\ref{fig_stepw}, for white noise. 

A similar trend is present here as with the flat profile: there is a steep change in spectral gradient between two regions of near-constant spectral gradient, as shown in Figure \ref{fig_flatgradsave}. The location of the steep gradient change in Figure \ref{fig_stepw} is independent of the noise source. An interesting feature is that the location of the steep spectral gradient change varies with plateau sizes. The location of the sharp change approximately follows the equation $a=-0.5b+1.5$ for lower plateau size $a$ and upper plateau size $b$. As the size of the upper plateau (b) increases, the size of the lower plateau (a) required for the sharp gradient change decreases but at a slower rate. This implies that the dimension of the lower plateau $a$ is more important in producing this steep change than the dimension of the upper plateau $b$. The marker on the vertical axis in this plot indicates the location of the steep change for the flat profile.

For a small lower plateau , there is near constant spectral gradient, similar to the results from the flat profile, Figure \protect\ref{fig_flatgrads}. In Figure \ref{fig_stepw} there is some variation present which is due to the stochastic element of these simulations. One would expect that averaging over several samples of the same noise colour would result in a smoother result, as with the flat profile, Figure \ref{fig_flatgradsave}.

As in the case of the flat profile, there is very little difference in the location of the sharp gradient change between the results for different noise colours. Any discrepancies are of the same order as those found for the flat profile (Section \ref{sec_flat}). For this reason, only the white noise results are presented. It can be seen that the trend for the flat profile (Figure \protect\ref{fig_flatgradsave}) continues the trend present for the step profile, Figure \ref{fig_stepw}.

\subsection{Varying the mid-chromospheric temperature jump}
\label{sec_tempjump}

\begin{figure}
\centering
\includegraphics[width=88mm]{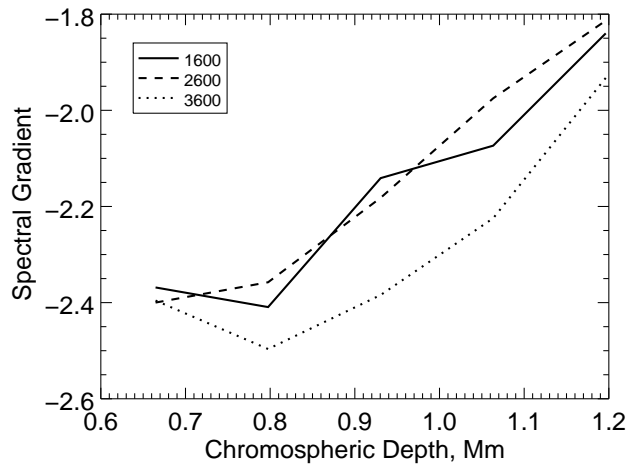}
\caption{Gradients of the spectra (Figure \protect\ref{fig_freq_reg}, region III) obtained with different temperature jumps at the mid chromosphere, i.e. with different values of $\Delta T$ as shown in Figure \protect\ref{fig_profiles}: 1600K, 2600K, 3600K. The chromospheric cavity size $a+b$ increases with $a = b$.}
\label{fig_temp_grad}
\end{figure}

A set of simulations was performed to investigate the importance of the temperature jump $\Delta T$ between the two uniform regions. For this, only a few chromospheric cavity sizes were tested, $0.6$ Mm $\leq a+b \leq 1.2$ Mm, and three temperature jumps were considered: $\Delta T=$ 1600, 2600 and 3600 K. The results are shown in Figure \protect\ref{fig_temp_grad}. There appears to be no significant difference between the temperature jumps, indicating that the variable $\Delta T$ is less important than the total chromospheric cavity size $a+b$. 

\section{Chromospheric seismology}
\label{sec_tempdiag}

The steep change in spectral gradient can be clearly seen in Figures \ref{fig_freq_reg} and \ref{fig_freq_reg_step}. Regions I and II in these figures are very similar. The diagnostic in this paper is demonstrated by considering region III of these two plots. In Figure \ref{fig_freq_reg} the chromospheric temperature profile leads to a shallow gradient, whereas the temperature configuration in Figure \ref{fig_freq_reg_step} leads to a steep gradient. One can clearly see the stark differences between these two plots for region III.  

The results obtained give rise to a new potentially useful diagnostic for chromospheric temperature configurations. Given the velocity spectra in the corona, it is possible to give upper and lower limits to the chromospheric cavity size. This diagnostic arises from the presence of the steep change in gradient present in all cases considered here, visible in Figures \ref{fig_flatgradsave} and \ref{fig_stepw}. For example, consider a case where the spectral gradient is steep. From Figure \ref{fig_stepw} it can be seen that the size of the chromospheric cavity is bounded above by the steep gradient change, which follows the line $a=0.5b +1.5$. This implies limits on the size of the lower plateau $0 \leq a \leq 1.5$ Mm and upper plateau $0 \leq b \leq 3$ Mm, subject to the constraint that $a +0.5b \leq 1.5$.

\section{Spectral shape and its height above the photosphere}
\label{sec_photospectra}

\begin{figure}
\centering
\includegraphics[width=88mm]{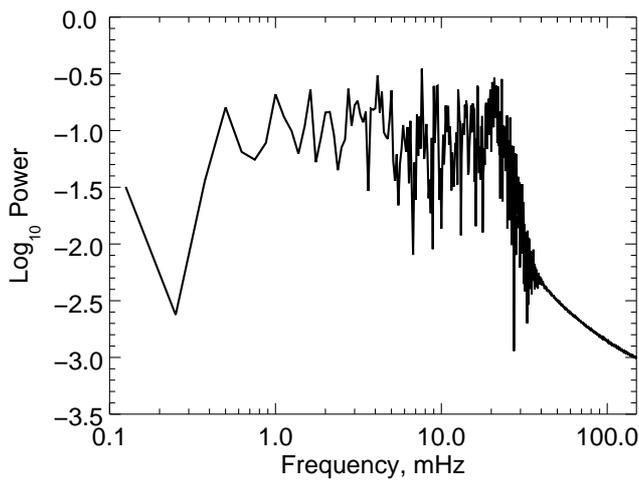}
\caption{Frequency spectrum sampled at the photosphere, i.e. at height 0.}
\label{fig_photo}
\end{figure}

In addition to the spectra in the corona, the photospheric spectra were also analysed for white noise. There is a reasonable flat spectra followed by a sudden drop at approximately 20 mHz, as shown in Figure \ref{fig_photo}. Again this sudden drop in power was found to be independent of the input wave resolution. A possible explanation for this is that the downwards propagating waves due to the resonator dwarf the driven upward waves in power at this point. Note that this does not mean the upward propagating waves are not present, merely that they are not obvious in the FFT since the waves propagating from the resonator are far more powerful. 

\begin{figure}[!h]
\centering
\includegraphics[width=88mm]{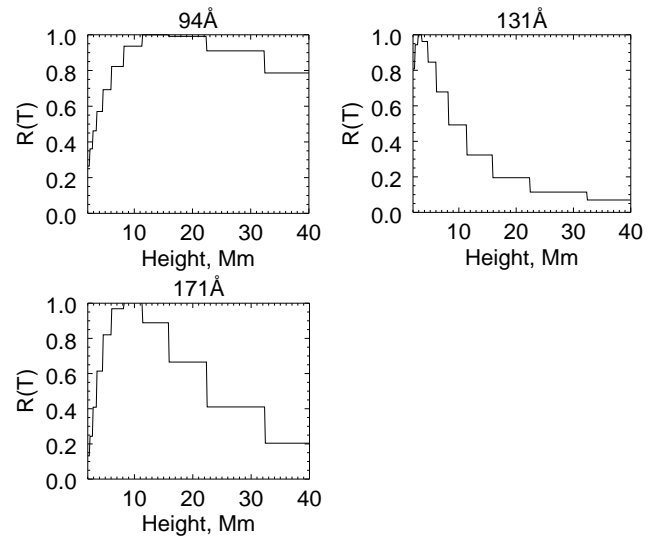}
\caption{Normalised response function $R(T)$ plotted against height using three SDO/AIA filters at time $t=0$. This is recalculated at each time iteration for the LOS integration to account for temperature variations. The corresponding initial temperature profile is shown in Figure \protect\ref{fig_temp}.}
\label{fig_respfuntemp}
\end{figure}

\begin{figure*}[!ht]
\sidecaption
\includegraphics[width=120mm]{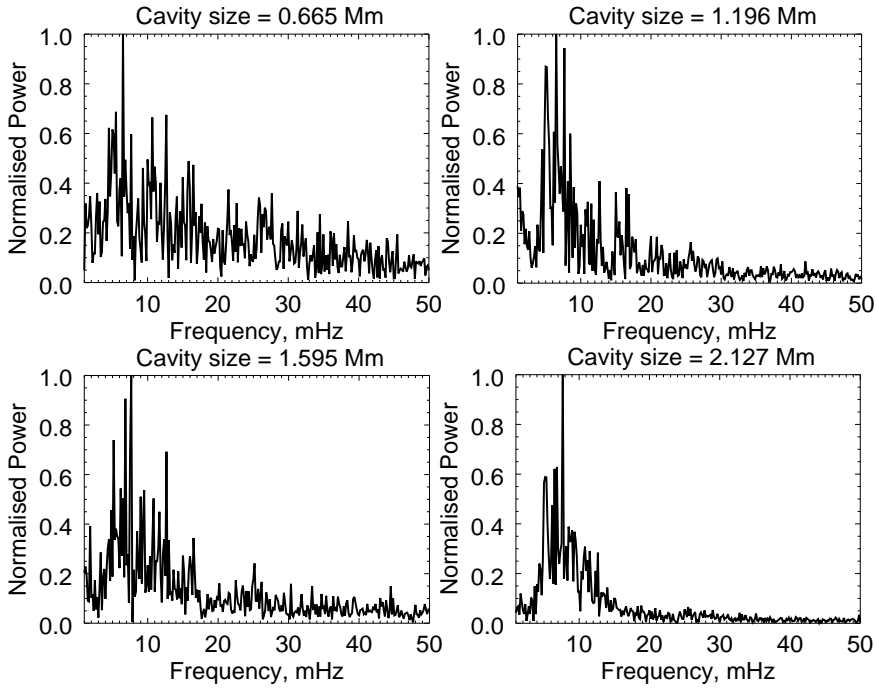}
\caption{Line of sight integration of the coronal velocity perturbations for 171 \AA~ for four different sizes of the chromospheric cavity. The power spectra are normalised to their respective maximum values for chromospheric cavity sizes (a+b in Figure \protect\ref{fig_profiles}) of the temperature profile given by 0.67, 1.20, 1.60 and 2.13 Mm respectively.}
\label{fig_LOSint}
\end{figure*}

It was also found that the spectra vary through the chromosphere. In particular, either side of a temperature jump i.e. the mid-chromospheric jump ($\Delta T$ in Figure \ref{fig_profiles}) and at the transition region, the higher frequencies in the spectra change behaviour. The upper spectral cut-off in Figure \ref{fig_photo} is very steep. However if a mid-chromosphere temperature jump is present, the signal in the upper chromosphere has a more gradual cut-off. This develops into the gradient and cut-off regions III and IV seen in Figures \ref{fig_freq_reg} and \ref{fig_freq_reg_step} above the transition region. Also, as expected, the acoustic cut-off frequency in Region I becomes more prominent towards the transition region. 

\section{Limitations of the observable coronal spectra}
\label{sec_LOS}
In order to investigate how the findings presented in this paper would be observed, LOS integration is performed. The line of sight integration is calculated using 

\begin{equation}
I=\int R(T) n_e ^2 dy
\label{eqn_LOS}
\end{equation} 
where $I$ is the resultant line intensity, $n_e$ is the number density, $R(T)$ is the response function for temperature $T$, and $y$ is the height. Note that the mass density $\rho$ is proportional to the number density $n_e$. This integration was performed from the transition region to the corona. This was performed numerically using the trapezium rule. The temporal resolution here is one sample every two seconds. This corresponds to a Nyquist frequency of 250 mHz. The response function is recalculated at each time step to account for temperature fluctuations. The integration was performed with a lower bound as the start of the transition region to remove the chromospheric foot-point. Only the flat profile driven by white noise has been considered. The line of sight integration was performed for chromospheric cavity sizes between 0.66 and 2.13 Mm.

\subsection{SDO/AIA response functions}
\label{sec_aialos}

The initial response function $R(T)$ as a function of height is shown in Figure \ref{fig_respfuntemp} for three SDO/AIA filters (94, 131, and 171 \AA). The SDO/AIA response functions were generated using the most recent version V6 available in ssw-idl \citep{Boerner2012}. LOS integration was performed on the three samples of white noise, varying the chromospheric cavity size between 5 and 2.13 Mm for the three SDO/AIA filters shown in Figure \ref{fig_respfuntemp}. The three SDO/AIA filters give similar results, hence we only discuss the behaviour for the 171 \AA\ channel.

Figure \ref{fig_LOSint} shows the resultant frequency spectra of the line intensity using the 171 \AA~ filter for chromospheric cavity sizes of 0.66, 1.20, 1.60 and 2.13 Mm. The spectra demonstrate the acoustic cut-off for frequencies less than 4 mHz, similar to the spectra obtained from a point source (Figures \ref{fig_freq_reg} and \ref{fig_freq_reg_step}, region I). As the chromospheric cavity size increases, the power of the underlying noise becomes less dominant and the highest frequency in the excited range decreases. In other words, the bandwidth containing the excited frequencies becomes narrower. For example, in Figure \ref{fig_LOSint} the excited range of frequencies for a chromospheric cavity size of 1.595 Mm is between approximately 4 and 18 mHz whereas for a chromospheric cavity size of 2.127 Mm the range is between 4 and 13 mHz.   

A line of the form $ 1/f^ \alpha $ can be fitted to the LOS spectra in the range 20-50 mHz, with $\alpha$ representing the log-log gradient. For the single point analysis there was a trend present where the gradient depended on the chromospheric cavity size. However only a weak trend is present in the gradients for the 94 and 171 \AA~ filters. No discernible trend was present for the 131 \AA~ filter. Variations in $\alpha$ between different samples of white noise are large with respect to the total change in $\alpha$. Therefore it would be difficult to accurately determine the chromospheric cavity size given $\alpha$.

\subsection{LOS integration vs single point analysis}

The LOS spectra are different to the frequency spectra in the single point analysis. The primary difference is the range of frequencies present in the broad peak. The lower limit of this region is the same, indicating that the acoustic cut-off frequency is detectable in both cases. However the upper limit of the broad peak is 10 mHz for the LOS spectra, but 17 mHz for the single-point spectra.

The LOS integration gives different results compared to the single point analysis. Possible explanations are as follows:

\begin{enumerate}
\item The integration of a sinusoidal function depends on the depth of the LOS integration and the periodicity of each individual component of the signal. As such, the power of the individual frequencies is modified non-uniformly by performing the LOS integration. This results in a different power distribution compared to the single point analysis. 
\item The spectral lines/filters have a finite width covering a temperature range. This is in contrast to the single point analysis that isolates a single temperature value. The same holds for the mass density, $\rho$.
\item The distribution of the frequency spectra changes with height in the transition region where the temperature gradients are large. As one moves into the corona, the variation in the shape of the frequency spectra decreases. Any spectral lines/filters including contributions from the transition region will be affected. The single point analysis results discussed in this paper are obtained in the corona. Therefore, the weighted transition region will enhance the differences between the single point analysis and the LOS integration. 
\end{enumerate}

\subsection{Angle of LOS integration}
In the 1.5D model the LOS integration is performed parallel to the magnetic field lines. Performing the integration perpendicular to the magnetic field will yield the single point result (since it is invariant in this direction). Integrating at an angle produces the same results as integrating parallel to the magnetic field line, since, even though the path length appears longer, it intercepts the same discrete values in the same order.

\subsection{Comparison with observations}

The frequency spectra obtained in Figure \ref{fig_LOSint} are very similar to observational papers as can be seen in \cite{Tian2014}; an excited range of frequencies is present between approximately 4.5 and 10 mHz \citep{Thomas1987,Rez2012,Tian2014}. For wider chromospheres, this is more distinct and the excited range is reasonably narrow. However for smaller chromospheres the LOS spectra are far more noisy and whilst there are a few strong peaks, the power does not decay rapidly, as can be seen in Figure \ref{fig_LOSint}. The bandwidth of the LOS frequency spectra can be used to estimate the chromospheric cavity size, as was discussed in Section \ref{sec_aialos}. For example the 4.5 to 10 mHz range observed by \cite{Thomas1987} would indicate a chromospheric cavity size of greater than 2.13 Mm.

\section{Conclusions}

In this paper, a numerical investigation of oscillations above a sunspot umbra was performed. Multiple simulations were performed where both the chromospheric temperature profile and perturbation noise were varied. The resultant coronal velocity signature was then analysed. Different colours of noise were found to have very little effect on the output spectra. However, a strong trend is present where the gradient in the frequency spectra of the coronal velocity perturbations varies with regards to chromospheric cavity size: as the chromosphere cavity increases in size, the spectral gradient becomes shallower. 

The coronal spectra can potentially be used as a diagnostic for the chromospheric cavity size. The sharp change in gradient present in Figures \ref{fig_flatgradsave} and \ref{fig_stepw} allows for limits to be estimated for the sizes of the lower and upper uniform temperature regions or plateaux ($a$ and $b$ in Figure \ref{fig_profiles}). These limits can be further reduced by estimating a maximum and minimum total chromospheric cavity size ($a+b$).  

The LOS spectra possess high power oscillations in the three-minute band, similar to those in observed spectra. It is found that the bandwidth of the excited frequencies become narrower as the chromospheric cavity size increases. This is a promising tool to estimate the chromospheric cavity size.

There was no clear relation present between the LOS spectral gradient and the chromospheric cavity size when integrating along a magnetic field line. The LOS integration along a magnetic field line masks the underlying complexity of the waves present in the single point analysis. This indicates that a very narrow response function is necessary to clearly observe a trend along a magnetic field line. Alternatively, observing the sunspot at an angle to the photosphere would yield a different view where the waves are not integrated along the magnetic field line. As such, one would expect the results to approach the results obtained from the single point analysis as the angle approaches 90$^{\circ}$ with respect to the magnetic field lines.

It is inherently difficult to measure the chromospheric cavity size directly. This work provides a solution by analysing the coronal waves above sunspots for different chromospheric cavity sizes via a numerical simulation. The single point results presented in this paper show that the gradient of the coronal frequency spectra is directly correlated with the chromospheric cavity size. This is equivalent to performing LOS integration perpendicular to the magnetic field line. When LOS integration is performed along the magnetic field line, it is shown that the bandwidth of excited frequencies varies with chromospheric cavity size. Thus, a novel diagnostic was presented for indirectly estimating the chromospheric cavity size above sunspot umbrae for the Sun and other stars. 

\begin{acknowledgements}
We thank the anonymous referee for constructive comments. The authors acknowledge IDL support provided by STFC. The computational work was supported by resources made available through Warwick University’s Centre for Scientific Computing and STFC HPC facilities. 
\end{acknowledgements}

\bibliographystyle{aa.bst}
\bibliography{Bibliography.bib}

\end{document}